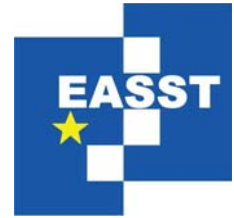

Proceedings of the
International Workshop on
Software Quality and Maintainability
(SQM 2014)

A Model-Based Approach to Impact Analysis Using Model Differencing

Klaus Müller, Bernhard Rumpe

15 pages

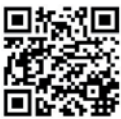

# A Model-Based Approach to Impact Analysis Using Model Differencing

**Klaus Müller**[1]**, Bernhard Rumpe**[2]

[1] mueller@se-rwth.de, [2] rumpe@se-rwth.de, http://www.se-rwth.de/
Software Engineering
RWTH Aachen University, Germany

**Abstract:** Impact analysis is concerned with the identification of consequences of changes and is therefore an important activity for software evolution. In model-based software development, models are core artifacts, which are often used to generate essential parts of a software system. Changes to a model can thus substantially affect different artifacts of a software system. In this paper, we propose a model-based approach to impact analysis, in which explicit impact rules can be specified in a domain specific language (DSL). These impact rules define consequences of designated UML class diagram changes on software artifacts and the need of dependent activities such as data evolution. The UML class diagram changes are identified automatically using model differencing. The advantage of using explicit impact rules is that they enable the formalization of knowledge about a product. By explicitly defining this knowledge, it is possible to create a checklist with hints about development steps that are (potentially) necessary to manage the evolution. To validate the feasibility of our approach, we provide results of a case study.

## 1 Introduction

A software system typically undergoes frequent modifications due to changing or new requirements or bug fixes. A major problem is that even small changes can have severe effects on a software system and that it is often hard to predict which parts of a software system are affected in what way by a change. Impact analysis approaches address this problem by identifying the potential consequences of a change [Boh95]. Existing work on impact analysis mainly focuses on the effects of code changes [Leh11].

In model-based software development, models are primary artifacts of development, which are typically transformed into concrete implementations [FR07]. Out of the multitude of different modeling techniques, UML class diagrams still constitute the most frequently used modeling technique of the UML [Rum11]. Although employing techniques such as code generation can drastically reduce the number and size of artifacts that have to be written manually [Rum12], it is usually still necessary to create and maintain a variety of artifacts manually, such as source code, configuration files or property files. As these artifacts have to be integrated into a generated infrastructure, they are sometimes heavily depending on the generated artifacts. For instance, the database schema might be generated based on a UML class diagram. If a handwritten source code file contains a SQL query which accesses this database, this source code file depends on the generated database schema and indirectly also on the UML class diagram. Changes in models can therefore have tremendous impact on handwritten artifacts.

In this paper, we propose a model-based approach to impact analysis. Explicit impact rules do not only embody which conditions have to be fulfilled by concrete UML class diagram changes to have an impact but they also contain a description of the actual impact. These conditions and consequences can be specified in a DSL, based on which an implementation is generated, which checks the specified conditions and outputs the defined impact.

The main motivation for explicitly specifying impact rules is that they allow leveraging known dependencies and characteristics of a product, as this knowledge can be formalized in an impact rule. By describing the impact of changes, it is possible to create a checklist with concrete and precise hints about further development steps that are (potentially) necessary to manage the evolution. This checklist can then be taken and worked through in the evolution process and thus provides a structured approach to adapt the system to the changes. With this approach, we do not only strive for identifying which artifacts are impacted by a change in what way, but also aim at advising the user of other dependent activities such as data evolution.

Our approach is based on the assumption that we are aware of the changes performed in UML class diagrams. These changes are identified automatically using model differencing. However, this automatic derivation of model differences cannot be done correctly in all cases. We tackle this problem by providing users a simple way to integrate knowledge of how selected model elements changed from one model (version) to the other. This knowledge is embodied by so-called user presettings [MR14]. A simple user presetting can, e.g., express that a specific model element should be regarded as renamed.

The paper is structured as follows: in Section 2, we describe the background of our work before we present some concrete examples for impact rules in Section 3. After this, we give a brief overview of our approach in Section 4. Subsequently, the identification of model differences is outlined in Section 5. Afterwards, our rule-based approach to impact analysis is described in Section 6. Section 7 shows the results of our case study before we present related work in Section 8. In Section 9 we discuss our approach. Finally, Section 10 summarizes the paper.

## 2 Background

Our work is motivated by software development projects of the company DSA [DSA] but also applicable to other situations. One of the most prominent DSA software solutions is PRODIS.Authoring [Aut] (hereinafter referred to as Authoring). It is a complex software system in which an essential part was generated out of UML class diagrams. The idea to develop this impact analysis approach arose out of discussions with Authoring developers who encouraged creating a checklist with concrete hints about (potential) changes induced by changes in UML class diagrams.

The motivation for creating such a checklist is that developers might not always remember all concrete development steps that have to be carried out after specific model changes - especially in a complex software system. This was reported to us in discussions with developers. But even if developers do not forget to perform these further activities, such a checklist will still be helpful as it guides the developers by listing which steps have to be performed next. This also comprises the identification and listing of potentially affected artifacts so that developers need not search for them themselves. By explicitly defining knowledge of a product in impact rules, we counteract a further problem: if developers leave a company without writing down their knowledge, their expertise might get lost, at least partly.

We are aware of the limitations of our approach, as we can only deal with impact induced by UML class diagram changes and not with the impact of code changes. However, our approach is not intended to replace all existing impact analysis approaches but is intended to be used as a complementary approach which focuses on the impacts of changes in UML class diagrams.

## 3 Illustrative examples

In this section, we present simplified examples for impact rules which could be used to support the evolution of Authoring. For each impact rule we also outline the consequences it has if the developers forget to perform the according development steps. In addition to that, we briefly discuss a concrete checklist that was created by some of these impact rules.

### 3.1 Impact rules

#### 3.1.1 XML migration analysis

Authoring provides the possibility to export and import XML data. However, XML data cannot always be imported directly, but sometimes they must be converted before. For this purpose, XML migration classes have to be implemented. Several model changes can necessitate this task and various special cases have to be taken into account. These different conditions are formalized in this impact rule, which notifies the users if it is necessary to create such a migration class. If possible, it also proposes ready-to-use migration classes.

If developers forget to implement XML migration classes, the customer might not be able to import XML data. As this feature is important, according to DSA, this would be really problematic from the customers perspective. But even if a bug is not reported by the customer but by the test team, the resulting costs should not be underestimated, as the cause of these bugs is not always detected immediately and in addition to fixing the bug, it is also necessary to retest it.

#### 3.1.2 SQL query analysis

At various points in Authoring, SQL queries are used to retrieve information from the database. As the database schema of Authoring is to a great extent generated out of UML class diagrams, class diagram changes can affect the database schema and lead to changes of, e.g., names of tables or columns. If manually written SQL queries refer to these table or column names, the SQL queries have to be changed as a consequence of the model change. This impact rule finds and reports SQL queries that need to be changed after having changed the model.

If a certain part of Authoring uses an invalid SQL query, the customer will usually not be able to use that part of the system properly. On the other hand, manually ensuring that all SQL queries are still valid is a laborious and error-prone task.

### 3.1.3 Object-relational mapping file analysis

Authoring uses an object-relational mapping (ORM) solution to map the object-oriented model to a relational database. The ORM files partly have to be changed manually after changing the model. This impact rule identifies and reports the elements which have to be added to the ORM files, that have to be changed in the ORM files and that should be deleted from the ORM files.

The access to the database can only work properly if the ORM files are configured correctly. Errors in the ORM files are therefore problematic for the whole system and all users. Although such errors should be detected at the latest by the test team, every support in this field is helpful to avoid bugfixing costs.

### 3.1.4 Property file analysis

For certain types of model elements in the UML class diagrams, the generator creates code which refers to keys in property files. The developers have to add descriptions for these keys, which are later shown in specific views in Authoring. This impact rule advises the developers of the necessity to add appropriate entries to the property files or to delete existing entries.

Missing entries in the property files do not necessarily lead to errors, but can also result in inappropriate strings being displayed in certain views in Authoring. This would decrease the usability of these views.

## 3.2 Checklist example

An excerpt of a shortened checklist for two fictive model differences is shown in Figure 1. The complete checklist is shown in [Che]. For each impact rule the checklist contains hints that describe which development steps have to be performed (line 3 && line $8-10$) and which model changes caused this (line 4 && line $9-10$). For instance, the checklist reports to the user that an entry should be added to the ORM file for the new class (line 3) or it can be seen which entries should be added to the property file (line $9-10$).

```
1 ORM file analysis:
2 =================
3 - Add entry to mapping file for new class.
4 (Causing model change: Added class 'de.test.ECU')
5
6 Property file analysis:
7 =======================
8 Add these entries to the property file core.properties:
9 -ECU (Causing model change: Added class 'de.test.ECU')
10 -ECUS (Causing model change: Added class 'de.test.ECU')
```

Figure 1: Shortened checklist example

## 4 Overview

Our approach to impact analysis is composed of two main steps: the identification of model differences and the creation of a checklist based on these differences. Both steps are outlined in the following enumeration. Furthermore, Figure 2 gives an overview of our approach. The next two sections describe both steps in more detail.

1. **Model differencing**: the model differencing component determines the differences between two models the user had to define as input. The users can verify the correctness of these differences and integrate user presettings, if they want to correct the reported differences. For the case that user presettings exist, these are taken into account in the model differencing process. The result of this step is a difference model which contains all differences.

2. **Checklist creation**: the difference model is traversed by the checklist generator to create the checklist. Each difference of the difference model is passed to the available impact rules. Each impact rule then analyzes the difference and creates a list of hints at further development steps, if the difference is considered relevant.

The whole process is implemented in a tool which creates a text file with the checklist as a result. The users can then take this checklist and incrementally tick the relevant steps off.

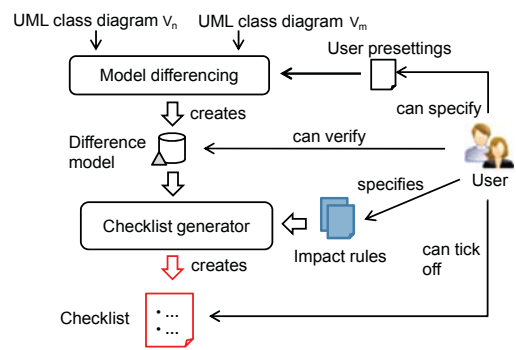

Figure 2: Overview of our approach to impact analysis

## 5 Model differencing process

In our approach, we use a slightly extended version of EMF Compare [EMF] to perform model differencing. As we did not develop a completely new approach to model differencing, we only give a brief introduction to the model differencing process applied in our approach.

### 5.1 Overview

In order to identify the differences between two models, three main steps are performed: at first the user has to define the models of which differences shall be identified. Normally two versions of the same model are used as we intend to identify what has changed within a certain model.

These models are then converted into Ecore files as EMF Compare is best suited to infer structural differences between Ecore files. To a great extent, this Ecore conversion is pretty straightforward as the standard elements of an UML class diagram can easily be mapped to Ecore. Due to this, we will not describe this conversion process in this paper.

The model differencing component finally builds the difference model containing all differences between the previously converted models. This difference model can contain all kinds of changes of Ecore elements such as additions, deletions, renamings or movements of model elements or other updates of model element properties such as the cardinality of an attribute.

### 5.2 Integration of user presettings

In general, a completely automatic approach to model differencing cannot infer the differences correctly in all cases [PMR13]. This can even hold for small changes like a renaming, as this could also be regarded as the deletion of a model element and the addition of another model element. Because of this problem, we allow users to integrate knowledge of how specific model elements changed from one model (version) to the other. In our approach, this is done using so-called user presettings [MR14]. As explained in [MR14], multiple user presetting instructions are provided, e.g., to express that a model element should be regarded as renamed or moved. Figure 3 shows an example for a rename presetting, indicating that the attribute `name` of the class `de.TroubleCd` was renamed into `newName`. Details on how the model differencing component processes user presettings are provided in [MR14]. It should be noted that user presettings only have to be provided if the identified model differences contain errors that need to be corrected. In all other cases, the users do not need to specify user presettings.

```
renamed "de.TroubleCd#name" to "newName";
```

Figure 3: User presetting instruction example

## 6 Rule-based impact analysis

In this section, we first give an overview of our approach. Then we elaborate on the specification of impact rules, before we present a concrete example.

### 6.1 Overview

In our approach to impact analysis, impact rules capture the consequences of UML class diagram changes. In an impact rule, the user is free to define what kind of change leads to what kind of impact - such as the modification of an existing artifact or the creation or deletion of a particular artifact. For that purpose the previously created difference model is traversed and each difference is passed consecutively to each impact rule. The impact rules then analyze the impact of the currently considered model difference.

To improve the comprehensibility of an impact rule, we provide a simple DSL. In this DSL, it can be specified which conditions have to be fulfilled by a model difference so that a certain checklist hint is created. The next subsection describes this aspect in more detail. Based on these specifications, Java implementations of the impact rules are generated. It is also possible to implement an impact rule directly in Java. Even though we recommend to use the DSL as far as possible, implementing the rules completely or to a great extent in Java is still useful as the knowledge about the impact is formulated explicitly and can be reused to support software evolution. One crucial aspect that has to be highlighted is that once an impact rule exists, this impact rule can be used subsequently every time the checklist generator is executed.

### 6.2 Impact rule specification

The basic structure of the impact rule DSL is illustrated in Figure 4. Vertical bars separate the different values which can be specified, a question mark following a parenthesis expresses that the corresponding part is optional and a star indicates that there can be zero or multiple elements of that type.

```
impactRule "<name>" {
  description = "..."
  (severity = minor|normal|critical)?
  (probability = low|medium|high)?
  (relevantFor = "...")?

  impact {
    (<condition part> => "<checklist hint>")*
  }
}
```

Figure 4: Basic structure of an impact rule

Each impact rule has a name (line 1) and a description (line 2) revealing what the impact rule is used for. Furthermore, a severity can be declared (line 3). This severity indicates how critical it is to perform the further reported development steps. The possible values for this are "minor", "normal" or "critical". Such an information can later be used by developers to define the chronology in which they perform the different activities. In addition to that, it can be stated how probable it is that the reported impact really holds (line 4). Possible values for this are "low", "medium" or "high". Moreover, each impact rule can refer to specific persons for which the hints created by the impact rule are potentially relevant (line 5). This information can be used to inform only selected developers about further development steps.

After this descriptive part, it is possible to define which conditions have to be fulfilled by a model difference to lead to the creation of the subsequently given checklist hint (line 8). Each impact rule can have zero or multiple blocks of condition parts and corresponding checklist hints. The complete condition part is treated like a boolean expression in Java and has to be a valid boolean expression due to this. In each condition part, multiple conditions can be combined such as boolean expressions can be combined in Java using logical operators. For this purpose, the DSL supports the logical operators && and ||. Furthermore, conditions can be negated using the operator ! and can be parenthesized. The meaning of these notations is the same as in Java. The condition part can refer to two types of conditions: predefined conditions and user-defined conditions. The meaning of both types of conditions is described in more detail after this paragraph. The user can also leave the impact block (line 7 - 9) empty. In this case, the condition checks and corresponding checklist hints have to be implemented by the user.

Predefined conditions are conditions which are part of the framework and which can be used as-is in every impact rule. In order to differentiate between predefined conditions and user-defined conditions, predefined conditions are introduced with the keyword `pc` followed by a dot. We provided them to simplify the specification of conditions in impact rules. For each type of model change which can be found in the model differencing activity (cf. Section 5.1), we offer one predefined condition, which checks whether the particular type of model change occurred. Hence, there exist predefined conditions like `deletedClass`, `movedAttribute` or `addedAssociation`. For example, the condition `addedAssociation` indicates, whether the currently considered model difference represents the addition of an association.

As we developed the DSL in the cooperation project with DSA, we added predefined conditions which are specific for Authoring. For instance, the UML class diagrams used for Authoring contain special stereotypes to mark persistent elements such as classes and attributes. The Authoring specific predefined conditions take advantage of the knowledge about this, so that they, e.g., check whether a persistent class was added or deleted. For these two given examples, the conditions `addedPersistentClass` and `deletedPersistentClass` were added.

We only created predefined conditions that are potentially relevant for all impact rules. As this might not always be sufficient to state conditions, we also allow the user to introduce self-defined condition names within this condition part. This way, the user can express that a further condition check has to be performed. As the impact rule generator cannot be aware of the implementation of these conditions, the user has to add the implementation for them manually.

To allow a flexible specification of impact rules, we also allow the user to insert place holders into the checklist hint. In doing so, the user can declare that further code has to be implemented which returns the concrete string that will be inserted at the particular place in the checklist hint. This can be necessary if the checklist hint cannot be defined completely in a static way.

The advantage of supporting user-defined conditions and place holders is that they allow users to formulate abstract conditions and the effects without having to precisely implement this immediately. This improves the comprehensibility of an impact rule as everybody can roughly understand which checks are performed without being aware of the implementation details.

Altogether, the user only has to extend the generated impact rule implementation, if at least one condition part refers to a user-defined condition or if at least one checklist hint contains a place holder. In order to integrate the manually written implementations and the generated impact rule implementations, we apply the generation gap pattern [Fow10]. Based on an impact rule specification, a base class is generated. The user can then add a manually written subclass. In this subclass, user-defined conditions have to be implemented or if place holders exist, code has to be implemented which returns the string that will be inserted into the checklist hint. For this purpose, the user has to adhere to specific naming conventions. Details on this are omitted here on purpose in order to focus on the concept and the ideas behind the approach.

So far, we only considered the impact of single model changes. Our approach is not limited to this but can also be used in situations in which the impact depends on multiple model changes. Although the impact rule DSL does not provide mechanisms to analyze multiple model changes, the user can add such checks in the handwritten impact rule implementation. This is possible as each impact rule has access to the whole difference model. Moreover, each impact rule also has access to the complete original and the complete changed class diagram. This can be helpful if the impact rule needs to obtain information from concrete elements of the class diagrams. Due to these possibilities, even sophisticated impact rules can be defined.

### 6.3 Impact rule example

Figure 5 shows an excerpt of the impact rule referring to changes in ORM files (cf. Section 3.1.3). It expresses that the ORM files have to be changed in two cases: if a persistent class was added (line 4) and that class is an active class (line 4) or if a persistent attribute was renamed (line 6). The condition `addedActiveClass` is a user-defined condition as this condition is not introduced by the keyword `pc`. The actual implementation of this condition has to be performed in the subclass of the generated base implementation. In what way the ORM files have to be changed is described as well (line 5 && line 6 − 7). This example also demonstrates how to insert place holders into the checklist hint part. In this case, the place holder has the name `ORMFileExcerpt` (line 7). Instead of this place holder, another string is inserted into the checklist. Which string exactly is integrated depends on the implementation of the subclass of the generated base implementation.

```
impactRule "ORM File Analysis" {
  description = "This rule checks ..."
  impact {
    pc.addedPersistentClass() && addedActiveClass() =>
      "Add entry to mapping file for new class."
    pc.renamedPersistentAttribute() => "Rename entry in
      mapping file. Excerpt from file: {ORMFileExcerpt}"
  }
}
```

Figure 5: Simplified impact rule for the ORM file analysis

## 7 Evaluation

### 7.1 Objective and research questions

The objective of our case study is to validate the applicability of our approach in a real-world software system. For this purpose, we derived the following three research questions.

- **RQ1**: is it feasible to use the approach for real-world models with respect to the execution time of the approach?

- **RQ2**: are the impact rules capable to identify the relevant development steps?

- **RQ3**: how do the developers of the software system assess our approach?

### 7.2 Design and subject of the study

We decided to apply our approach on Authoring, as it is a complex real-world software system applying model-based software development. At first, we calculated the model differences between 20 versions of three larger UML class diagrams used for Authoring. These class diagrams contained roughly between 100 and 300 classes. All successive versions were compared pairwise, resulting in $(20-1) \cdot 3 = 57$ model comparisons. These model updates occurred in about one year of development. In total, 243 model elements were added, 12 model elements were deleted, 23 model elements were moved and 12 model elements were renamed.

In order to answer the first research question, we measured the execution times of applying the approach pairwise on the UML class diagrams, as described before. To validate the applicability for big models, we also applied our approach on a UML class diagram containing 4000 classes. On that one, we performed 500 change operations, covering all types of UML class diagram changes that are relevant for the impact rules, and measured the execution time. These experiments were performed on a Dell Latitude E3620 (8GB RAM; Intel i7-2620M 2.7 Ghz).

To answer the second research question, we implemented selected impact rules introduced in Section 3, compared the old versions of the UML class diagrams, as explained before, and assessed the hints generated by the impact rules.

To answer the third research question, we passed five representative checklists to five developers. Three of these developers were experienced developers and the other two knew the system well but had to carry out development tasks rather seldom. In addition to that, we passed questionnaires to the developers and asked them to answer the following questions openly. Moreover, they could write down further open feedback on the approach.

- **DQ1**: how much do the checklists ease performing the further development steps?

- **DQ2**: how do you assess the costs of working through the checklists?

- **DQ3**: what are the reasons for you to (not) use the checklists in the future?

In the following, the results for each research question are presented in different subsections.

### 7.3 Results and discussion

#### 7.3.1 Research Question RQ1

The complete execution time for performing all 57 pairwise model comparisons together with the checklist creation was 37 seconds. This is the average execution time that resulted from ten executions. For the UML class diagram with 4000 classes, the complete execution time was 15 seconds on average. Out of these measurements we conclude that it is feasible to apply the approach for real-world models and also in situations with a high amount of model changes.

### 7.3.2 Research Question RQ2

At first, we present the results of the analysis concerning the XML migration classes. Our impact rule reported 15 different migration classes. These comprised all nine migration classes that have been created by the developers. Please note that from these, eight migration classes were not implemented immediately but later in the development process as bugs were reported by the test team. From the six other migration classes, one migration class was a false positive, three have to be implemented for the next release and two have been missed in the past. Not only has the impact rule listed the migration classes that had to be implemented, but in 12 cases concrete migration classes were proposed and from these 10 could have been used directly for the real system without having to change them.

Next, we dissected the results of the ORM analysis impact rule. Our search for bugs in this area could only find one bug. A further analysis has shown that our approach referred to all changes that had been integrated into the ORM files for the already known 57 model updates. Although the probability that developers forget to perform this development step seems to be very low, our impact rule can assist the developers by reporting necessary changes. This reduces the required time to perform the changes.

Furthermore, we analyzed the results of the property file impact rule. We could only find one bug indicating that there were missing entries in the property files. The impact rule reported 275 concrete entries that should have been added to the property files. Out of these, actually 254 were added. The others were not added and are currently still missing.

These results show that our approach was capable to refer to the performed development steps and that it also revealed development steps that were forgotten in the past. Hence, the approach is already helpful for the current development process.

### 7.3.3 Research Question RQ3

In question DQ1, four users reported that the checklists do ease performing the further development steps and one developer said that the checklists would be helpful if they would not contain so many details. For instance, one developer stated: "I think the information shown there is extremely valuable.". Another one wrote down: "The probability that further development steps are forgotten decreases. They animate to work in a more structured way.".

The same developer that criticized the size of the checklists in the first question, also answered in question DQ2 accordingly: "The checklists are bigger than necessary. Their size acted as a deterrent at first sight.". The other four users answered that the costs of working through the checklists are acceptable, although one user also mentioned that certain details could be omitted. For example one developer answered: "I expect the checklists to speed up the development.". Another developer pointed out that the level of detail of the checklists were perfectly right.

The analysis of the replies to question DQ3 showed that four developers would like to use the checklists in future and one developer would use them if they would be shorter. Out of the four developers, one said: "I would always use the checklists. In that way, you cannot forget to perform certain steps, no matter how big or small the changes are.".

In the remaining open feedback part two developers proposed to use a different format for the checklist, to make it easier to tick off what was already done.

All developers appreciated the idea of having a checklist which can be ticked off. However, the feedback from the developers also revealed that there are varying opinions on the proper size of a checklist. Due to this we plan to extend our approach by creating one detailed checklist with additional explanations and one short checklist containing only the really needed information.

### 7.4 Threats to validity

The major threat to the internal validity is that other changes require to perform the according development steps and not only the identified model changes. Anyhow, we analyzed the effects of the model changes isolated from other changes to prevent other changes from influencing the results.

The major threat to the external validity is that we applied the approach only on a single, but real-world, software system. However, the initial situation that an essential part of the system is generated based on UML class diagrams and that changes in UML class diagrams can necessitate performing further development steps, is no speciality of this software system. In addition, there were no further special characteristics which made this software system particularly suited for our approach. Due to this, we are convinced that our approach is also applicable to other (complex) software systems applying model-based software development. Furthermore, we are aware of the fact that we cannot draw general conclusions from the results of five questionnaires. Anyway, we passed the checklists to those developers that are the most affected by the generated checklists and that are therefore best able to assess the checklists. Due to this, we regard the feedback of these developers as particularly relevant.

## 8 Related Work

Existing work on impact analysis mainly focuses on the effects of code changes [Leh11]. These approaches usually use static or dynamic program analysis techniques to identify affected code parts. In addition, there are various other techniques, such as information retrieval techniques, approaches mining software repositories or traceability approaches. A further class of approaches is based on using explicit impact rules that express the consequences of a change. In the following, we will present some rule-based approaches in more detail, as our approach is also rule-based.

The works of Sun et. al [SLT$^+$10], Chaumun et al. [CKKL02] and Queille et. al [QVWM94] all present code-based analysis techniques assuming that the impact of a change depends on the type of change and on the type of relationship between the artifacts. In these cases, impact rules refer to potentially affected parts of a program. In addition to that, in the work of Queille et. al [QVWM94] propagation rules denote what kind of modifications have to be performed on affected objects. In contrast to these works, our approach analyzes model changes and their impact. Moreover, we do not mainly focus on identifying impacted entities of a program but also support users by creating concrete hints at further activities.

Briand et. al [BLO03] aims at identifying the impact of UML model changes on other UML model elements. At first, the changes between two versions of UML models are automatically inferred. After that, the impacts of the changes are computed based on rules which are defined using the Object Constraint Language. These reveal which UML model elements should be regarded as affected. Unlike the work of Briand et. al, we do not primarily aim at identifying impacted model elements but want to be able to refer to `any` kind of impact.

Lehnert et. al [LFR13] address the problem that changes in artifacts (e.g. UML models or Java source code) can induce changes in a variety of other artifacts. At first, explicit rules determine between which types of artifacts which type of dependency link should be created. On top of this, they perform change propagation by using explicit impact propagation rules. These rules basically express that the impact depends on the dependency relation between artifacts and the type of change applied on them. The work of Lehnert et. al is more general than ours, as it determines the impact of changes of different kinds of artifacts. In return, we can capture dependent activities that do not concern existing artifacts. In addition, we can more easily create accurate hints at further development steps to manage the evolution.

## 9 Discussion and Future Directions

In this section, we discuss topics associated with our work, limitations and future work directions.

In our approach, impact rules embody consequences of syntactic changes in UML class diagrams. Thereby we cannot deal with the impact of code changes or changes in other types of models. Despite this, our approach is generally not limited to UML class diagrams. As outlined in Section 5, we perform a conversion from UML class diagrams into Ecore files at first. After that, we create the difference model for the resulting Ecore files. As a result, we can easily extend our approach to be able to capture changes of other types of models by adding a converter from the corresponding language to Ecore.

The next issue is related to false positives reported by impact rules. If checklists contain many false hints on development steps, developers might decide to not use the checklists at all. To mitigate this problem, we provide the possibility to declare the probability of the reported impact. Hence, developers can start working off the impact which is regarded as most probable.

One further issue concerns conflict management within impact rules. The approach is not able to detect whether hints created by impact rules contradict each other. Instead, we assume that the users specifying impact rules ensure this on their own.

The last issue we want to discuss deals with the cost-benefit ratio of our approach. The key question that has to be answered before implementing an impact rule is whether the implementation effort is worth it compared to the benefits gained. This cannot be answered generally but has to be answered individually for each impact rule. The key aspect that has to be taken into account when assessing the usefulness of an impact rule is that once it is implemented, it can be executed repeatedly (at any time) and it always checks in the same way whether a change has an impact and what kind of impact. This can have a tremendous positive impact on software quality, as the probability that a developer forgets to perform the according development steps decreases significantly. Beyond that, the checklists can make it easier for the developers to perform the development steps, as they do not have to analyze on their own, e.g., which model change affects the ORM files or in what way property files have to be changed. One further aspect which is relevant for a cost-benefit analysis is whether new versions of the system are planned to be released regularly or not. The more often new versions of a system are released, the more often can the impact rules assist the developers in their development steps. Thus, the implementation effort can amortize in the course of time, as the developers are supported continuously and after each change they are relieved of having to perform the particular checks such as validating whether existing SQL queries are affected by model changes. Finally, even though the impact rule implementation has to be done manually, this allows for creating precise hints that are tailored to the developers. An automatic approach will not accomplish this in this way. Please note that we must not state numbers on how much time is saved by means of the impact rules evaluated in Section 7. Furthermore, we must not name concrete numbers on the effort that had to be spent for implementing these impact rules.

Currently, we assume that developers work through the checklists and perform the listed changes, but we do not control this. For future work, we plan to investigate for which checklist hints we can check automatically whether the user really implemented the particular change. As already stated in Section 7.3.3, we also plan to extend our approach by creating one detailed checklist with additional explanations and one short checklist containing only the really needed information. In addition, we plan to improve the format of the checklist to make it easier for users to tick development steps off. One possible way to improve this would be to connect the developed tooling to a issue tracking system such as JIRA [Jir] and to create issues for the further (potential) development steps.

## 10 Conclusion

In this paper, we have proposed an approach to impact analysis in which explicit impact rules capture the consequences of UML class diagram changes on other artifacts and determine the need of dependent activities such as data evolution. The analyzed UML class diagrams typically describe two versions of the system under development, and differences are identified automatically using model differencing. By explicitly formulating consequences of changes in impact rules, we are capable of creating a checklist with accurate hints concerning development steps that are (potentially) necessary to manage the evolution.

This approach is mainly feasible for software systems in which an essential part of the system is generated, but also a lot is written by hand and sometimes heavily depends on the generated code. In these – rather common – cases, assistance using checklists for updating handwritten code or performing further development steps is pretty helpful.